\documentstyle[12pt]{article}
\begin{document}
\parindent 1.3cm
\thispagestyle{empty}   % to suppress the page number on the first page
\vspace*{-3cm}
\noindent

\def\arccot{\mathop{\rm arccot}\nolimits}
\def\sd{\strut\displaystyle}

\begin{obeylines}
\begin{flushright}
NORDITA-93/13 N,P
UAB-FT-299/92
UG-FT-27/92
\end{flushright}
\end{obeylines}

\vspace{2cm}

\begin{center}
\begin{bf}
\noindent
SEMILEPTONIC $\pi$ AND K DECAYS AND THE CHIRAL ANOMALY AT ONE-LOOP
\end{bf}
  \vspace{1.5cm}\\
Ll. AMETLLER
\vspace{0.1cm}\\
Departament F{\'\i}sica i Enginyeria Nuclear, FIB, UPC,\\
08028 Barcelona, Spain\\
\vspace{0.5cm}
J. BIJNENS
  \vspace{0.1cm}\\
NORDITA, Blegdamsvej 17,\\DK-2100 Copenhagen \O, Denmark\\
\vspace{0.5cm}
A. BRAMON
   \vspace{0.1cm}\\
Grup de F\'\i sica Te\`orica, Universitat Aut\`onoma de Barcelona,\\
08193 Bellaterra (Barcelona), Spain\\
   \vspace{0.5cm}
and \\
\vspace{0.5cm}
F. CORNET
\vspace{0.1cm}\\
Departamento de F\'\i sica Te\'orica y del Cosmos,\\
Universidad de Granada, 18071 Granada, Spain\\
   \vspace{2.2cm}

{\bf ABSTRACT}
\end{center}

We calculate the $O(p^6)$ corrections to the anomalous form factors
appearing in $\pi^+$, $K^+ \to e^+ \nu \gamma,\ \mu^+\nu\gamma$
and $K_{l4}$ decays
in Chiral Perturbation Theory. The relevant dimension $6$ terms of
the lagrangian are evaluated assuming their saturation by the vector
meson contribution.

\newpage

Chiral Perturbation Theory (ChPT) is believed to furnish one of the most
appropriate and accurate descriptions of the low energy interactions of
pseudoscalar mesons. This includes not only their strong interactions,
but also the electroweak semileptonic
ones, whose dynamics is completely
fixed once the corresponding gauge bosons have been introduced through
the usual covariant derivative. The perturbative or series expansion in
ChPT is made in terms of the particle four-momenta and pseudoscalar masses,
which are therefore required to be small compared to a typical
symmetry breaking scale assumed to be around 1 GeV \cite{GL} \cite{L}.
For all these reasons, the semileptonic and radiative weak decays
of pions and kaons are expected to be most accurately
described in ChPT. Indeed, one has to deal exclusively with well fixed (no
free-parameter) Lagrangians and the energies involved belong to
the lowest region of the spectrum,
with the masses of the two lightest hadrons as an upper bound. Our purpose
is to compute and discuss the anomalous form factor of the
 $\pi, K \to \gamma l\nu$ and $K_{l4}$ decays. In this sense, the present
note is an extension of the recent work of one of the authors on the
non-anomalous sector of $K_{l4}$ decays, to which we refer for further
details \cite{Hans} (see also Refs. \cite{Rig,dafne,rad}).\par
The octet of pseudoscalar mesons $M = \sum_a \pi^a\lambda^a/\sqrt{2}$
appears in ChPT through the $3 \times 3$ matrix
\begin{equation}
\Sigma\equiv \exp{{2i\over f}M},
\end{equation}
where f can be identified with the pion decay constant
$f_\pi=f=132 \rm{MeV}$ ($=f_K,$ at lowest order). As previously stated,
the electroweak gauge fields
$A_\mu$ and $W_\mu$ are introduced via the covariant derivative
\begin{equation}
\begin{array}{l}
D_\mu= \partial_\mu\Sigma +i L_\mu\Sigma-i\Sigma R_\mu;\quad \cr
\ \cr
L_\mu=e A_\mu Q
   -\sd{g\over \sqrt{2}}\pmatrix{0 & \cos\theta W_\mu^+ & \sin\theta W_\mu^+
                                                                           \cr
                                        \cos\theta W_\mu^- & 0 & 0 \cr
                                        \sin\theta W_\mu^- & 0 & 0 \cr }; \quad
R_\mu=e A_\mu Q,
\end{array}
\end{equation}
where $Q$ is the quark charge matrix in units of $e=g \sin\theta_W>0$,
and $\theta$ and $\theta_W$ stand for the Cabibbo and Weinberg angles. The
final lepton pair (for definiteness we will always refer to $l^+\nu$ coming
{} a $W^+$) appears in the leptonic current $l_\mu$ when substituting
\begin{equation}
W_\mu \quad \to \quad {g\over 2\sqrt{2}M_W^2} l_\mu \equiv
{g\over 2\sqrt{2}M_W^2} \bar u_\nu \gamma_\mu (1 - \gamma_5) v_e,
\end{equation}
with $G_F/\sqrt{2}=g^2/8M_W^2$.

The Lagrangian of ChPT has an anomalous and a normal (or non-anomalous)
sector. At lowest order ($p^2$ or second order), the latter is fully given
by
\begin{equation}
L_2={1\over 8}f^2~tr[D_\mu\Sigma D^\mu\Sigma^\dagger+ \chi\Sigma^\dagger+
\Sigma \chi^\dagger],
\end{equation}
where the last $\chi$-terms contain the SU(3)-breaking quark mass matrix.
The anomalous sector begins at fourth ($p^4$) order with the Wess-Zumino
term $L_{WZ}$ \cite{WZ} containing pieces (among others) with zero, one and
two gauge boson
fields. The five pseudoscalar (no-boson) term relevant for our purposes is
given by
\begin{equation}
L_{WZ}^{(0)}= -{2\over 5\pi^2 f^5}\epsilon^{\mu\nu\alpha\beta}~
tr(M\partial_\mu M\partial_\nu M\partial_\alpha M\partial_\beta M).
\end{equation}
The terms with one and two gauge bosons have the form
\begin{equation}
\begin{array}{ll}
L_{WZ}^{(1)}= & -\sd{1\over{16\pi^2}}\epsilon^{\mu\nu\alpha\beta}~tr
(\Sigma \partial_\mu\Sigma^\dagger\partial_\nu\Sigma\partial_\alpha
\Sigma^\dagger L_\beta
-\Sigma^\dagger\partial_\mu\Sigma\partial_\nu\Sigma^\dagger
\partial_\alpha\Sigma R_\beta)
\cr\ \cr
L_{WZ}^{(2)}= & -\sd{i\over{16\pi^2}}\epsilon^{\mu\nu\alpha\beta} tr
\big[\partial_\mu\Sigma^\dagger\partial_\nu L_\alpha \Sigma R_\beta
-\partial_\mu\Sigma\partial_\nu R_\alpha \Sigma^\dagger L_\beta \cr
{}~~~& ~~~~ \cr
{}~~~~&
+\Sigma\partial_\mu\Sigma^\dagger (L_\nu\partial_\alpha L_\beta + \partial_\nu
L_\alpha
L_\beta)
\big],
\end{array}
\end{equation}
where only the terms needed for the processes discussed here are given.

With these Lagrangians one immediately obtains the tree-level amplitude
for the anomalous part of our processes. For the radiative $\pi$ and $K$ weak
decays one has
\begin{equation}
\label{treerad}
\begin{array}{l}
A^{an}(\pi\to l\nu\gamma)= \sd{e G_F \cos \theta\over \sqrt{2} m_\pi}
F_V^\pi \epsilon^{\mu\nu\alpha\beta}l_\mu q_\nu\varepsilon_\alpha k_\beta
\cr  \  \cr
A^{an}(K\to l\nu\gamma)= \sd{e G_F \sin \theta\over \sqrt{2} m_\pi}
F_V^K \epsilon^{\mu\nu\alpha\beta}l_\mu q_\nu\varepsilon_\alpha k_\beta,
\end{array}
\end{equation}
with $F_V^\pi=m_\pi/4\pi^2f$ and $F_V^K=m_K/4\pi^2f$.
For the different $K_{l4}$ amplitudes one similarly gets\footnote{The
normalization of the $K_{l4}$ form factors used here is that of
Ref. \cite{Hans}. Conversion to those of Ref. \cite{dafne} can be done
by comparing the respective lowest order expressions.}
\begin{equation}
\label{kl4tree}
A^{an}(K\to \pi^i\pi^j l\nu)= -\sqrt{2} G_F \sin\theta {1\over m_K^3}
H^{ij} \epsilon^{\mu\nu\alpha\beta} l_\mu q_\nu p_\alpha^i p_\beta^j,
\end{equation}
where the indexes i, j=+,--,0 refer to the pion charge states and $H^{00}=0,
H^{+-}=H^{-0}/\sqrt{2}=-m_K^3/2\pi^2f^3$. Among these three isospin rotated
amplitudes one can verify the relation $A^{-0}=\sqrt{2}(A^{+-}-A^{00})$ valid
not only at this tree-level but at all orders. In the amplitudes above, $K$,
$p^{i,j}$, $k$ and $q$ stand for the four-momenta of the decaying
pseudoscalar, final pions, photon (with polarization $\varepsilon$) and
lepton pair, respectively, i.e., $K=q+k$ or $K=p^i+p^j+q$.

One-loop corrections to the above tree-level amplitudes proceed through
diagrams similar to those in Refs. \cite{Hans,ZfP},
as well as from the usual renormalization
of the pseudoscalar wave-functions and decay constants. For the two
radiative decays we have found

\begin{equation}
\label{piradloop}
\begin{array}{ll}
A^{an}(\pi\to & l\nu \gamma) =
                \sd{e G_F \cos \theta\over 4\sqrt{2}\pi^2 f_\pi}
\epsilon^{\mu\nu\alpha\beta} l_\mu q_\nu \varepsilon_\alpha k_\beta \Big\{
 1 + \sd{1\over 16\pi^2f^2}\Big[
                            {2\over 3}\lambda(k^2+q^2) \cr
{}~~ & ~~~ \cr
{}~~~ &              \sd{-4 m_\pi^2 \ln{m_\pi^2\over \mu^2}
                           -4 m_K^2   \ln{m_K^2  \over \mu^2} }
  + 4I(q^2,m_\pi^2,m_\pi^2)+4I(k^2,m_K^2,m_K^2)\Big]+C_\pi\Big\},
\end{array}
\end{equation}

\begin{equation}
\label{kradloop}
\begin{array}{ll}
A^{an}(K\to & l\nu \gamma) =  \sd{e G_F \sin \theta\over 4\sqrt{2}\pi^2 f_K}
\epsilon^{\mu\nu\alpha\beta} l_\mu q_\nu \varepsilon_\alpha k_\beta \Big\{
 1 + \sd{1\over 16\pi^2f^2}\Big[
{5\over 3}\lambda (m_K^2-m_\pi^2) \cr
{}~~ & ~~~ \cr
{}~~~ &
       +\sd{2\over 3}\lambda(k^2+q^2)-{7\over 2} m_\pi^2
                                                     \ln{m_\pi^2\over \mu^2}
                         -3 m_K^2   \ln{m_K^2  \over \mu^2}
                        -{3\over 2} m_\eta^2   \ln{m_\eta^2  \over \mu^2}\cr
{}~~~ & ~~~ \cr
{}~~~ &
  + 4I(k^2,m_\pi^2,m_\pi^2)+2I(q^2,m_K^2,m_\pi^2)
+2I(q^2,m_K^2,m_\eta^2)
\Big]+ C_K\Big\}.
\end{array}
\end{equation}
Similarly, for the various $K_{l4}$ amplitudes we obtain
\begin{equation}
\label{kpml4loop}
\begin{array}{ll}
A^{an}(K^+ \to & \pi^+\pi^- l^+\nu)=
\sd{G_F \sin\theta\over \sqrt{2}\pi^2 f_K f_\pi^2}
\epsilon^{\mu\nu\alpha\beta} l_\mu q_\nu p_\alpha^+ p_\beta^-\cr
{}~~ & ~~~ \cr
{}~~~ &
\Big\{
 1 + \sd{1\over 16\pi^2f^2}\Big[
                       {7\over 6}\lambda(m_{\pi}^2-m_K^2)
+{1\over 3}\lambda[(p^+ + p^-)^2+(p^-+q)^2+ 2 q^2]\cr
{}~~~ & ~~~ \cr
{}~~~ &
                     -\sd{11\over 2} m_\pi^2 \ln{m_\pi^2\over \mu^2}
                          -5 m_K^2   \ln{m_K^2  \over \mu^2}
                          -{3\over 2} m_\eta^2  \ln{m_\eta^2  \over \mu^2} \cr
{}~~~ & ~~~ \cr
{}~~~ &
  + 2I((p^+ +p^-)^2,m_\pi^2,m_\pi^2)+I((p^+ +p^-)^2,m_K^2,m_K^2)
  + 2I((p^- +q)^2,m_K^2,m_\pi^2) \cr
{}~~~ & ~~~ \cr
{}~~~ &
  +  I((p^- +q)^2,m_K^2,m_\eta^2)
  + 3I(q^2,m_K^2,m_\pi^2)+3I(q^2,m_K^2,m_\eta^2)
\Big] + C_{+-}\Big\},
\end{array}
\end{equation}

\begin{equation}
\label{k00l4loop}
\begin{array}{ll}
A^{an}(K^+ \to & \pi^0\pi^0 l^+\nu)=
\sd{G_F \sin\theta\over \sqrt{2}\pi^2 f_K f_\pi^2}
\epsilon^{\mu\nu\alpha\beta} l_\mu q_\nu p_\alpha p'_\beta\cr
{}~~ & ~~~ \cr
{}~~~ &
\Big\{
 0 + \sd{1\over 16\pi^2f^2}\Big[
  -{1\over 3}\lambda[(q+p)^2 - (q+p')^2] \cr
{}~~~ & ~~~ \cr
{}~~~ &
  - I((q+p)^2,m_K^2,m_\pi^2)-{1\over 2}I((q+p)^2,m_K^2,m_\eta^2)\cr
{}~~~ & ~~~ \cr
{}~~~ &
  +\sd I((q+p')^2,m_K^2,m_\pi^2)+{1\over 2}I((q+p')^2,m_K^2,m_\eta^2)
\Big]+ C_{00}\Big\}.
\end{array}
\end{equation}
The $A^{00}$ amplitude is the even part of the $A^{+-}$ amplitude
under the interchange $p^i \leftrightarrow p^j$ as required
by isospin. The $A^{-0}$ amplitudes are the odd part of $A^{+-}$ or can be
determined by the relation discussed above.

All the four one-loop amplitudes above have been expressed in a similar
form. The first term inside the brackets (vanishing in the last amplitude)
is essentially the tree-level term quoted in eqs.(\ref{treerad}),
(\ref{kl4tree}). Notice however
that the latter eqs. always contain the unbroken decay constant f, whereas in
the one-loop corrected versions one has the corresponding $f_\pi$ or $f_K$
decay constant. Next, one has two types of divergent terms proportional to
$\lambda=1/\epsilon +1+\ln{4\pi}-\gamma_E$ coming from our dimensional
regularization scheme, which also introduces the subtraction mass $\mu$. These
terms will be discussed later on together with the contributions from the
higher order effective Lagrangian, the $C_i$.
 Finally, one has the so called chiral
logs incorporating explicit logarithmic terms plus the function $I$ to which
we now turn.

The defining equation for the function $I$ and its relation to the simpler
functions $A$ and $B$ is given by
\begin{equation}
\begin{array}{l}
I(k^2,m_1^2,m_2^2)  \equiv \cr
{}~~~~ \cr
\sd\int_0^1dx[m_1^2-(m_1^2-m_2^2) x -x(1-x)k^2]
\ln\sd{[m_1^2-(m_1^2-m_2^2) x -x(1-x)k^2]\over \mu^2 }\cr
{}~~~~~~~ \cr
\hskip 2truecm =16\pi^2\Big[\sd{m_1^2-m_2^2+k^2\over 6 k^2}iA(m_1^2) +
              \sd{-m_1^2+m_2^2+k^2\over 6 k^2}iA(m_2^2) \cr
{}~~~~~~ \cr
\hskip 2truecm -\sd{(m_1^2-m_2^2-k^2)^2-4k^2m_2^2\over 6 k^2}
                                                        iB(k^2,m_1^2,m_2^2)
        \Big]
-\sd{1\over 3}(m_1^2+m_2^2) +\sd{1\over 9}k^2,
\end{array}
\end{equation}
where
\begin{equation}
iA(m^2)={m^2 \over 16\pi^2}\ln{m^2 \over \mu^2}
\end{equation}
and
\begin{equation}
iB(k^2,m_1^2,m_2^2)= \frac{-1}{16\pi^2}\Bigg\{ 1 -\frac{1}{2}\log
\frac{m_1^2 m_2^2}{\mu^4}+ \frac{m_2^2 -m_1^2}{2 k^2}\log\frac{m_1^2}{m_2^2}
-\frac{1}{k^2} {u_+} {u_-} \log\frac{{u_+} + {u_-}}{{u_+} - {u_-}}\Bigg\}\ ,
\end{equation}
with $u_\pm = \sqrt{k^2 - (m_1 \pm m_2)^2}$,
coincide with the first two equations of
Appendix A in ref. \cite{Hans} with the opposite signs. In particular, for
$m_1=m_2$ and small values of $k^2$ one has
\begin{equation}
I(k^2,m^2,m^2)=m^2 \ln{m^2\over \mu^2}-{k^2 \over 6}(1+\ln{m^2 \over
\mu^2})+{\cal O}(k^4)
\end{equation}
This expansion allows for a clear check of our expressions: making all the
various squared four-momenta in the $I$-integrals tend to zero and using the
last expansion implies the vanishing of all the chiral logs, in agreement
with the well-known non-renormalization of the anomaly in the soft limit.

As previously mentioned, there are two types of divergent
$\lambda$-terms. Some are proportional to the SU(3)-breaking mass-difference
$m_K^2-m_\pi^2$ and have been derived using the Gell-Mann-Okubo relation
$(3m_8^2=4m_K^2-m_\pi^2)$ to eliminate the $\eta$-mass. The other type of
divergent terms contain the various combinations of squared four-momenta in
the final state. Again, both types of terms vanish in the soft limit for
masses and four-momenta. Also, and this is another test of our calculation,
these two types of terms could be expected on the basis of the analysis made in
ref. \cite{ZfP}. In this paper, all the divergent terms appearing at the
one loop level in the anomalous sector were identified and classified in
two groups: the SU(3)-breaking terms proportional to the M-meson squared
mass differences and terms containing combinations of squared four-momenta.
It is easy to identify in each group of this general expression  each one
of the divergent terms obtained in eqs.
(\ref{piradloop}--\ref{k00l4loop}).
Obviously, the presence of
these divergent terms requires the introduction of the corresponding
counterterms in the anomalous part of the
Lagrangian at order $p^6$. Examples of
these counterterms were already given in refs. \cite{PRL},\cite{eta}. Their
infinite part cancels the $\lambda$-terms in eqs.
(\ref{piradloop}--\ref{k00l4loop}),
which are simply
substituted by the remaining finite part of the counterterms,
$C_i^r$. The value of
these finite contributions from the counterterms to our processes is not
fixed in ChPT. It has to be deduced from data fitting (as was done in ref.
\cite{GL,dafne} for the counterterms of the non-anomalous Lagrangian at order
$p^4$) or, alternatively, from the hypothesis of resonance saturation of
the counterterms. \par
This hypothesis of resonance saturation was already proposed by Gasser and
Leutwyler in their original papers on ChPT \cite{GL} and has been fully
exploited and confirmed by Ecker et al. \cite{EGPdR} and other authors. In
our case involving photons, lepton currents and  pseudoscalar pairs,
one expects the vector-mesons to play the essential role. These are easily
incorporated in the chiral Lagrangian as ``massive" Yang-Mills fields
\cite{Meissner} or in the so called ``hidden symmetry" scheme of Bando et
al.\cite{Bando}. For our processes, all the relevant dynamics is generated
by the vector-vector-pseudoscalar (VVM) coupling discussed in refs.
\cite{ZfP} and \cite{eta}, where details can be found. The contribution of
vector-mesons to the first type of (SU(3)-breaking) counterterms is found
to vanish in all our amplitudes. By contrast, it is found to give a
definite finite contribution to the other type (four-momenta dependent)
counterterms. For the two radiative weak decays, vector-meson dominance of
the counterterms implies that in eqs. (\ref{piradloop}) and
(\ref{kradloop}) one has
\begin{equation}
C_\pi^r = C_K^r =  {k^2\over M_V^2} + {q^2\over M_V^2} + ...
\end{equation}
where the dots stand for contributions of heavier resonances than the vector
mesons.
As expected, this $k^2$ and $q^2$ dependence coincides
with the one obtained in ref. \cite{PRL} when dealing with the purely
anomalous $\pi^0, \eta$ couplings to two off-mass-shell photons. For $K_{l4}$
decays, the vector meson contributions give in eq.
(\ref{kpml4loop})
\begin{equation}
C_{+-}^r =  {3\over 2} {q^2\over M_V^2} + {3\over 4} {(p^++p^-)^2\over M_V^2}
+ {3\over 4} {(q+p^-)^2\over M_V^2} + ...
\end{equation}
where the dots again refer to contributions from heavier resonances.
These $M_V^2$ dependent terms
are the analogues of those found in ref. \cite{eta} when studying the $\eta
\to \pi^+\pi^-\gamma$ decays. Similarly, in eq. (\ref{k00l4loop})
we obtain
$C_{00}^r = 3(q+p)^2/8M_V^2-3(q+p')^2/8M_V^2+...$

Having calculated the vector-meson contribution to our
one-loop amplitudes we are able
to present our predictions.
These are summarized in Figs. 1 and 2 for the decays $\pi^+ \to e^+ \nu
\gamma$ and $K^+ \to e^+ \nu \gamma$, respectively. The quantities plotted,
$F^P$, are the respective form factors normalized to $1$ for vanishing
masses and four-momenta:
\begin{equation}
F^P = \sd{4 \pi^2 f_P \over m_P} F^P_V.
\end{equation}
These form factors are also valid for the decays with muons in the final state.
These are, however, suppressed by phase space compared to the decays involving
electrons.
In both cases the $O(p^6)$ corrections introduce a dependence of the
form factors with the $e^+ \nu$ and photon invariant masses. The corrections
obtained for the $\pi$ radiative decay are very small (smaller than $5 \%$)
all over the allowed kinematical range (Fig. 1). The case for
$K^+ \to e^+ \nu \gamma$ is more interesting because the corrections are
much larger (Fig. 2). They can become as large as $\sim 50 \%$ at
the edge of
the allowed kinematical range. There the number of events is, however,
very small. In any case, in the kinematical regions where the number of
events is larger the corrections can be as large as $10$ -- $20 \%$.
Finally, in Fig. 3 we present our results for the $H^{+-}$ form factor
as a function of the ($\pi^+ \pi^-$) invariant mass, and we compare
them to the experimental data \cite{Rosselet}. We have evaluated the form
factor taking a lepton invariant mass
of 100 $MeV$ and a $\cos\theta_\pi$ of 0.2. We keep these values
fixed because this was how the experiment was analyzed and these
values represent the peak of the observed contributions.
The lowest order prediction with
$f_K = f_\pi$, $H^{+-} = - 2.67$, is given by the dotted line. The predicted
value of $H^{+-}$ increases when one takes the lowest order expression
with $f_K = 1.2 f_\pi$ (dashed line), as is sometimes done in current
algebra. However, when the effects of the
loop corrections and the higher order terms in the lagrangian are
consistently taken into account, the previous increase is almost canceled.
So, the total prediction turns out to be very similar to the lowest order
one with $f_K = f_\pi$.

The question of the saturation of the counterterms only by vector mesons
deserves some comments. It is certainly very reasonable for the second
(four-momenta dependent) type of counterterms, but the vanishing
contribution generated by this saturation to the other type of counterterms
looks more suspicious. To generate a non-vanishing contribution one should
invoke the presence of scalar resonances with known couplings to
$MW\gamma$ states. Experimentally nothing is known about this type
of couplings; even the nature
and properties of the scalar mesons are highly controversial.
There are two indications that the effects of the scalar mesons are rather
small. In the non-anomalous sector (see \cite{EGPdR})
their couplings tend to be
smaller than those of the vector mesons and, in addition, their contributions
are suppressed by an extra factor of $M_S^2/M_V^2 \approx  0.6 $.
The other
indication suggesting that such scalar mesons contributions
are small follows from our previous work on the
$\pi^0, \eta$ coupling to virtual photons \cite{PRL}, where
similar contributions should be present. The good agreement obtained
between the data and the predictions from ChPT without scalar meson
contributions can be interpreted as an indication of the
smallness of the latter.

In summary, we have discussed de $O(p^6)$ corrections to the anomalous
form factors in $\pi^+$, $K^+ \to e^+ \nu \gamma$ and $K_{l4}$ decays.
These include the corrections from loop diagrams
and the ones from higher dimension terms in the lagrangian.
The former can be exactly calculated in terms of the known parameters
of the lagrangian. In order to estimate the latter we have assumed that
they are saturated by the vector meson contribution. In
$\pi^+ \to e^+ \nu \gamma$ and $K^+ \to \pi^+ \pi^- e^+ \nu$ the corrections
obtained are very small. The case of $K_{l4}$ is particularly interesting
because there is a cancellation between the effects of $SU(3)$-breaking
correction in the meson decay constants and the contribution from the loop
diagrams, higher dimension terms and wave function renormalization.
Unfortunately, the errors in the available experimental data on this decay
are too large to extract any conclusion. Finally, the corrections obtained
for the decay $K^+ \to  e^+ \nu \gamma$ are larger. In all the cases the
main contribution to the corrections is given by the $O(p^6)$ tree level
terms.

\newpage
\begin{center}
{\bf Figure Captions}
\end{center}

Fig. 1: $F^\pi$ form factor at $O(p^6)$ (normalized to one for
vanishing masses and four momenta) for $\pi^+ \to l^+ \nu \gamma$
as a function of the photon squared invariant mass, $k^2$, and
the $l^+ \nu$ squared invariant mass, $q^2$.

Fig. 2: $F^K$ form factor at $O(p^6)$ (normalized to one for
vanishing masses and four momenta) for $K^+ \to l^+ \nu \gamma$
as a function of the photon squared invariant mass, $k^2$, and
the $l^+ \nu$ squared invariant mass, $q^2$.

Fig. 3: $~H^{+-}$ form factor at $O(p^6)$ for
$K^+ \to \pi^+ \pi^- l^+ \nu$ as a function of the $\pi^+ \pi^-$
invariant mass $E_{\pi \pi}$.


\begin{thebibliography}{99}

\bibitem{GL}
J. Gasser and H. Leutwyler, Nucl.Phys.B250 (1985) 465.
\bibitem{L}
H. Leutwyler, Proceedings of the XXVI International Conference on High
Energy Physics, Dallas 1992.
\bibitem{Hans}
J. Bijnens, Nucl. Phys. B337 (1990) 635.
\bibitem{Rig}
C. Riggenbach et al., Phys. Rev. D43 (1991) 127.
\bibitem{dafne}
J. Bijnens, G. Ecker and J. Gasser, Semileptonic kaon decays in
Chiral Perturbation Theory, in ''DA$\Phi$NE Physics Handbook,'' eds. L. Maiani,
G. Pancheri and N. Paver, INFN-Frascati, 1992.
\bibitem{rad}
J. Bijnens, G. Ecker and J. Gasser, Radiative semileptonic kaon decays,
preprint CERN-TH 6625/92, BUTP-92/38.
\bibitem{WZ}
J. Wess and B. Zumino,Phys. Lett. B37 (1971) 95,\\
E. Witten, Nucl. Phys. B233 (1983) 422.
\bibitem{ZfP}
J. Bijnens, A. Bramon and F. Cornet, Z. Phys. C 46 (1990) 599; see also
D. Issler, SLAC report SLAC-PUB-4943 (1989).
\bibitem{PRL}
J. Bijnens, A. Bramon and F. Cornet, Phys. Rev. Lett. 61 (1988) 1453;
J. Donoghue and D. Wyler, Nucl. Phys. B316 (1989) 289
\bibitem{eta}
J. Bijnens, A. Bramon and F. Cornet, Phys. Lett. B237 (1990) 488;
Ll. Ametller et al., in Proceedings of the Daphne Workshop; ed. G.
Pancheri, INFN; Frascati 1991.
\bibitem{EGPdR}
G. Ecker, J. Gasser, A. Pich and E. de Rafael, Nucl. Phys. B321 (1989) 311.
See also J. F. Donoghue, C. Ramirez and G. Valencia, Phys. Rev. D39 (1989)
1947.
\bibitem{Meissner}
U.-G. Meissner, Phys. Rep. 161 (1988) 213.
\bibitem{Bando}
M. Bando et al., Phys. Rep. 164 (1988) 217.
\bibitem{Rosselet}
L. Rosselet et al., Phys. Rev. D15 (1977) 574.
%\bibitem

\end{thebibliography}
\end{document}